\documentclass[twocolumn, twocolappendix]{aastex7}

\usepackage{amsmath}
\usepackage{amssymb}
\usepackage{bm}
\usepackage{xcolor}
\usepackage{dsfont}
\usepackage{booktabs}

\received{1 December 2025}
\revised{18 February 2026}
\accepted{19 February 2026}

\newcommand{\parens}[1]{\left(#1\right)}
\newcommand{\brackets}[1]{\left[#1\right]}

\graphicspath{{./}{figs/}}
\begin{document}

\title{\textit{Chandra} Proper Motions and Milliarcsecond Astrometry of Nineteen Pulsars}

\author[0000-0002-6401-778X]{Jack T. Dinsmore}
\email{jtd@stanford.edu}
\affiliation{Department of Physics, Stanford University, Stanford CA 94305}
\affiliation{Kavli Institute for Particle Astrophysics and Cosmology, Stanford University, Stanford CA 94305}
\email{jtd@stanford.edu}
\author[0000-0001-6711-3286]{Roger W. Romani}
\affiliation{Department of Physics, Stanford University, Stanford CA 94305}
\affiliation{Kavli Institute for Particle Astrophysics and Cosmology, Stanford University, Stanford CA 94305}
\email{rwr@astro.stanford.edu}

\begin{abstract}
We present X-ray proper motion (PM) measurements of 19 pulsars using new and archival data from the \textit{Chandra} X-ray Observatory, including pulsar wind trails and X-ray filaments. Precise X-ray PMs are often limited by uncertainties in aligning observations to a common reference frame. Our analysis uses unresolved X-ray flux from stars in the \textit{Gaia} catalog in addition to X-ray bright point sources for alignment, improving uncertainties. We obtain absolute positions referenced to \textit{Gaia} with typical astrometric precision $\sim10$\,mas and PM statistical uncertainties down to 1.3 mas yr$^{-1}$, the most precise X-ray PM achieved to date. With our improved frame alignment, PM accuracies are now limited by the pulsar flux in most cases. These results reveal a new X-ray filament and illuminate the wind nebula structures and origins of several of these pulsars.
\end{abstract}


\section{Introduction} \label{sec:intro}

Pulsar proper motions (PMs) are a valuable tool for understanding pulsar wind nebulae (PWNe) and for connecting pulsars to possible birth sites. For example, supersonic pulsars can form bow shock PWNe and leave X-ray radiating trails of turbulent shocked wind as they travel through the interstellar medium (ISM). These trails' morphology depends strongly on the pulsar velocity \citep{bykov2017pulsar, barkov20193d, olmi2019full}. Pulsar X-ray filaments are long, narrow X-ray structures misaligned with the pulsar PM, a rare phenomenon observed for very high velocity pulsars \citep{bandiera2008on, dinsmore2024catalog}. PMs also help constrain birth kicks and can provide kinematic ages for individual pulsars by pointing away from potential birth sites.

Very long baseline interferometry (VLBI) can provide precise PMs for radio-loud pulsars. Millisecond pulsars can be stable enough for high precision timing to deliver accurate PMs. But for young $\gamma$-ray discovered pulsars, with no radio detection and high timing noise, PMs must be constrained from their high energy emission. The \textit{Chandra} X-ray Observatory's exceptionally tight on-axis point-spread function (PSF) has enabled several X-ray-only pulsar PM measurements \citep{van2012chandra, de2013fast, halpern2015proper, auchettl2015xray, de2021psr, long2022proper, holland2025proper}. \textit{Chandra}'s $\sim 1''$ absolute pointing accuracy must by corrected by aligning observations to a common coordinate system using field point sources (PSs). Only after alignment does the pulsar's relative shift provide the PM. While X-ray counts from field PSs usually outnumbers those from the pulsar, \textit{Chandra}'s PSF widens rapidly with off-axis angle, so alignment uncertainty often limits PM precision, especially in sparse fields.

Prospects have substantially improved for X-ray PM measurements in recent years due to the accurate astrometry of billions of optical PSs provided by the \textit{Gaia} DR3 catalog \citep{gaia2023gaia}. Accurate \textit{Gaia} positions improve frame alignment using X-ray bright stars \citep[e.g.][]{holland2025proper}, but most \textit{Gaia}-associated X-ray counts come from X-ray-faint PSs that are only detectable when many sources are stacked. Using \textit{Gaia} astrometry as a guide, these faint counts can improve alignment accuracy.

\begin{figure*}
    \centering
    \includegraphics[width=0.95\linewidth]{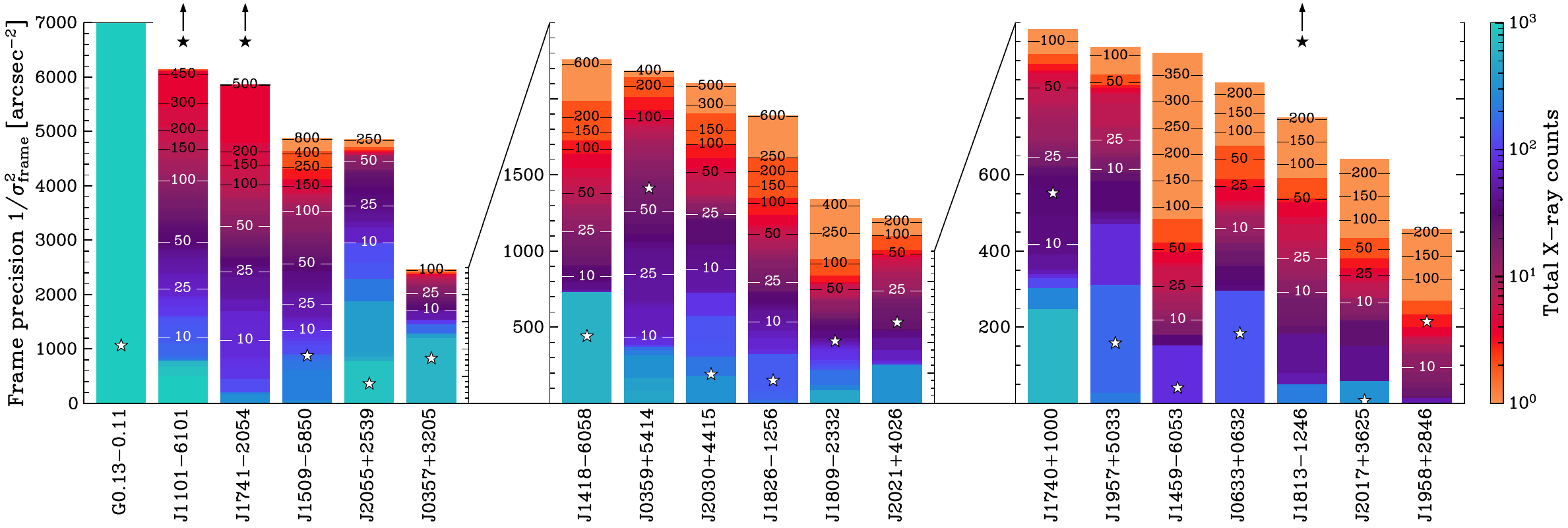}
    \caption{Estimated frame alignment precision for our pulsar fields. Histogram bar numbers indicate the cumulative number of PSs included in the fit, and redder colors correspond to X-ray faint, \textit{Gaia}-selected PSs. The pulsar astrometric precision is marked with a star. Whenever this star lies above the blue portion of the bars, \textit{Gaia} PSs are major contributors to the final PM precision.}
    \label{fig:field}
\end{figure*}

This work presents \textit{Chandra} PM measurements from new and archival observations of 19 pulsars, using \textit{Gaia}-associated counts to aid alignment. \S\ref{sec:methods} introduces our fit method and estimates the improvement expected from using X-ray-faint \textit{Gaia} sources. When possible, we use the PM to associate the pulsar with possible birth sites, estimate its age, and search for faint X-ray trails. These methods are discussed in \S\ref{sec:physics}. Results are discussed in \S\ref{sec:results}. We estimate the (small) systematic uncertainties in \S\ref{sec:systematics}, and conclude with \S\ref{sec:conclusion}.

\section{Methods} \label{sec:methods}
We first summarize the basic data preperation (\S\ref{sec:data}) and then note the precision gained from using \textit{Gaia} PSs (\S\ref{sec:estimation}). \S\ref{sec:fit-method} describes the statistical method for the actual PM fit.

\subsection{Data Preparation}
\label{sec:data}

This analysis uses all available \textit{Chandra} observations of each pulsar (appendix \ref{app:obs}), restricted to $0.5-7$ keV events and reprocessed with standard \texttt{ciao} tools. For several pulsars, we collected new observations to increase the temporal baseline and deliver more precise PMs. After merging all available exposures with the alignment provided by the observatory, we select bright X-ray PSs for use in precision frame registration. We choose only isolated PSs within 8$'$ of the pulsar and far from the deep observations' chip gaps. Depending on the field star density, $30-100$ X-ray PSs are selected. We also select all \textit{Gaia} DR3 PSs within 4' of the pulsar, excluding those in regions with X-ray nebulosity, which contributes another $70-760$ PSs.

An accurate PSF model is crucial for PM measurement when comparing \textit{Chandra} observations with different orientations or pointings. We start by simulating PSF ``libraries''---20$''$-spaced grids of PSFs each generated with the \texttt{simulate\_psf} \texttt{ciao} command on a grid with 16 spatial bins per \textit{Chandra} pixel. This command uses the ray-tracing MARX PSF model. We then linearly interpolate the library PSFs to the PS position. Since the \textit{Chandra} PSF is energy-dependent, we create separate ``hard'' and ``soft'' PSF libraries for both the ACIS-I and ACIS-S imagers. The ``hard,'' active galactic nuclei (AGN)-like spectrum is a power law with $dN_\gamma/dE \propto E^{-2}$, and the ``soft,'' star-like spectrum is thermal Bremsstrahlung with temperature $kT = 100$ eV. We label X-ray bright PSs as hard or soft by measuring their high- to low-energy count ratio $c_X = N(E > 2.5\, \mathrm{keV})/N(E < 2.5\, \mathrm{keV})$. Those with $c_X > 0.42$ are marked as hard (This threshold was chosen to best separate the presumably stellar X-ray PSs in DR3 from the presumably AGN X-ray PSs not in DR3). We treat all X-ray faint \textit{Gaia} PSs as soft since these provide insufficient counts to measure $c_X$.

\subsection{Uncertainty Improvement from \textit{Gaia} PSs}
\label{sec:estimation}

To illustrate the precision gained by using \textit{Gaia} PSs, Fig.~\ref{fig:field} reports the ``frame precision'' $1 / \sigma_\mathrm{frame}^2$ available from all the field PSs. $1 / \sigma_\mathrm{frame}^2$ is a sum over all counts from field PSs, background subtracted and weighted by the inverse PSF area since tighter PSFs better constrain the pointing.\footnote{We derive our frame precision metric as follows: A PS with $N_s$ counts detected above background should have position uncertainty of $\sigma_s \approx r_s / \sqrt{N_s}$, where $r_s$ is the root mean square event displacement calculated from our PSF images}. The uncertainty on the frame position $\sigma_\mathrm{frame}$ should be an inverse quadrature sum of $\sigma_s$: $1/\sigma_\mathrm{frame}^2 = \sum_s 1/\sigma_s^2 = \sum_s N_s / r_s^2$. Colors indicate the precision that would be achieved if only PSs exceeding a minimum number of counts were used. If the pulsar is used to assist in frame alignment, then the pulsar precision (marked with a star) is added to the total frame precision and alignment is further improved. Our analysis, described in the next section, uses both the X-ray faint stars and the pulsar to achieve optimal uncertainties.

\textit{Gaia} stars are a substantial aid whenever the red-yellow colors contribute a major fraction of the bar. And whenever the pulsar astrometry precision star lies above the blue region of the error bars, \textit{Gaia} stars are dominant contributors to the final PM accuracy.

This uncertainty metric is useful for a quick characterization of astrometric gain as in Fig.~\ref{fig:field}. But for the main results we estimate uncertainties more accurately using the likelihood itself.

\subsection{Fit Method}
\label{sec:fit-method}
Each PS $s$ has a position $\bm x_s$ and PM $\bm \mu_s$. For \textit{Gaia} PSs, $\bm x_s$ and sometimes $\bm \mu_s$ are known at the \textit{Gaia} epoch of 2016. We assume all unknown PMs are zero. This leaves the positions of the non-\textit{Gaia} PSs and the pulsar position and PM as free parameters, together with the offset $\bm \delta_f$ of each frame $f$. For long exposures, the aim point may drift linearly over time. We therefore introduce another fit parameter $\dot {\bm \delta}_f$ describing this drift rate for observations longer than 50 ks. We do not consider errors in the reported telescope roll, which should be subdominant to pointing variations.

The traditional method registers frames sequentially and then fits for the pulsar motion. However, a statistically optimal analysis fits all parameters simultaneously, which allows the pulsar's (often bright) signal to aid frame alignment while still accounting for its linear PM trend. We fit simultaneously using the maximum likelihood method. For each PS $s$ in frame $f$, we select the set of events $i$ within radius $R_s$ of the source.\footnote{To ensure we include every event possibly associated with the source, we choose $R_s = 2'' + 2r_s$, where $r_s$ is the local PSF radius. For the pulsar, we expand the radius to $R_s = 4'' + 2r_s$ so that pulsars with large PM do not leave the region of interest (ROI).} We perform our fit by maximizing the total likelihood $L$, which is a product of the likelihood $L_{i,s,f}$ for each event $i$ associated with PS $s$ and detected in frame $f$. That is, $L = \prod_s \prod_f \prod_i L_{i,s,f}$. The event likelihoods are
\begin{equation}
    L_{i,f,s} = \brackets{f_sP_{s,f}(\bm y_i - \bm y_{i,s,f}) + \frac{1-f_s}{\pi R_s^2}}C_{\mathrm{exp},f,s}^{-1},
    \label{eqn:like}
\end{equation}
 where $\bm y_i$ is the position of event $i$, $t_i$ is the arrival time relative to the \textit{Gaia} epoch, and $\bm y_{i,s,f} = (\bm x_s - \bm \delta_f) + t_i (\bm \mu_s-\dot {\bm \delta}_f)$ is the predicted position of the PS. The first term describes the fraction of ROI events $f_s$ that arrive from the PS. Here, $P_{s,f}$ is the PSF, cubic spline-interpolated from the appropriate library for the position of PS $s$ and the pointing of frame $f$, and normalized so that $\int d\bm y\, P_{s,f}(\bm y)=1$. The second term represents background events, which we assume are spatially uniform. Chip gaps, as encoded in the telescope exposure map, decrease both the background and PS probabilities at a given point equally (up to exposure map energy dependence, which induces minor corrections and is therefore neglected). The position of the likelihood maximum is therefore unchanged, so we do not need to multiply $P_{s,f}$ by the exposure map. However, the model should avoid over-predicting events in low-exposure regions. This is ensured by $C_{\mathrm{exp},f,s}$, which
 is set such that $L_{i,f,s}$ is normalized over event positions distributed by the exposure map. $C_{\mathrm{exp},f,s}$ depends on the predicted PS position, which we can approximate with a first-order Taylor expansion using pre-computed values of $C_{\mathrm{exp},f,s}$ and its gradient for each frame and PS. The exposure map was generated at 2.3\,keV with the \texttt{ciao} tool \texttt{fluximage}.

\begin{table*}
    \centering
    \hspace{-2.2cm}
    \begin{tabular}{lccccccc} \hline \hline 
        Name & RA & Dec & $\mu_x$  & $\mu_y$  & $d$ & $v_\perp$ & $Z$ \\ 
        & & & [mas yr$^{-1}$] & [mas yr$^{-1}$] & [kpc] & [km s$^{-1}$] & \\ \hline 
        G0.13$-$0.11 & 17:46:21.575(1) & $-$28:52:56.65(2) & $-1(2)$ & $-1(2)$ & 8 & $70(90)$ & 0.3 \\ 
        J0357+3205 & 03:57:52.279(2) & 32:05:21.23(3) & $-93(5)$ & $123(5)$ & $\sim0.6$ & $\sim420(10)$ & 31.5 \\ 
        J0359+5414 & 03:58:53.732(3) & 54:13:13.89(3) & $5(4)$ & $17(5)$ & $\sim3.9$ & $\sim330(80)$ & 3.5 \\ 
        J0633+0632 & 06:33:44.137(2) & 06:32:30.41(3) & $-14(5)$ & $20(5)$ & $\sim1.0$ & $\sim110(20)$ & 4.9 \\ 
        J1101$-$6101 & 11:01:44.932(1) & $-$61:01:39.27(1) & $-25(2)$ & $-21(2)$ & 6.3 & $990(50)$ & 20.8 \\ 
        J1418$-$6058 & 14:18:42.667(4) & $-$60:58:03.04(3) & $16(3)$ & $3(3)$ & $\sim1.4$ & $\sim110(20)$ & 4.6 \\ 
        J1459$-$6053 & 14:59:30.237(8) & $-$60:53:20.18(6) & $20(7)$ & $-1(8)$ & $\sim1.4$ & $\sim140(50)$ & 2.2 \\ 
        J1509$-$5850 & 15:09:27.150(3) & $-$58:50:55.94(2) & $10(4)$ & $17(4)$ & 3.4 & $310(60)$ & 4.7 \\ 
        J1740+1000 & 17:40:25.958(3) & 10:00:06.51(4) & $21(6)$ & $41(5)$ & 1.2 & $270(30)$ & 8.1 \\ 
        J1741$-$2054 & 17:41:57.258(2) & $-$20:54:13.00(2) & $-84(7)$ & $-84(7)$ & 0.27 & $154(9)$ & 17.7 \\ 
        J1809$-$2332 & 18:09:50.243(2) & $-$23:32:22.72(3) & $12(3)$ & $-6(3)$ & $\sim0.6$ & $\sim40(9)$ & 4.2 \\ 
        J1813$-$1246 & 18:13:23.757(2) & $-$12:46:00.66(2) & $-20(4)$ & $26(4)$ & $\sim1.6$ & $\sim260(30)$ & 7.7 \\ 
        J1826$-$1256 & 18:26:8.532(2) & $-$12:56:34.54(4) & $-9(4)$ & $3(4)$ & $\sim1.1$ & $\sim50(20)$ & 2.0 \\ 
        J1957+5033 & 19:57:38.373(4) & 50:33:21.11(4) & $-48(7)$ & $50(7)$ & $\sim0.8$ & $\sim270(30)$ & 10.1 \\ 
        J1958+2846 & 19:58:46.898(3) & 28:46:02.77(4) & $-12(5)$ & $1(5)$ & $\sim1.2$ & $\sim70(30)$ & 1.7 \\ 
        J2017+3625 & 20:17:55.822(9) & 36:25:07.7(1) & $-30(20)$ & $40(20)$ & $\sim0.6$ & $\sim150(50)$ & 2.5 \\ 
        J2021+4026 & 20:21:30.760(3) & 40:26:46.09(3) & $51(6)$ & $13(5)$ & $\sim0.3$ & $\sim84(9)$ & 9.4 \\ 
        J2030+4415 & 20:30:51.405(3) & 44:15:38.98(3) & $20(10)$ & $50(10)$ & 0.75 & $180(40)$ & 4.1 \\ 
        J2055+2539 & 20:55:48.959(2) & 25:39:58.64(3) & $50(9)$ & $-90(8)$ & $\sim0.6$ & $\sim280(20)$ & 11.8 \\ 
        \hline \hline 
    \end{tabular}
    \caption{Positions (J2000 at the \textit{Gaia} epoch of 2016) and proper motions for the pulsars considered in this work. Axes are right ascension and declination. The corresponding plane-of-sky velocity and significance of the PM detection $Z$ are also listed. Distances marked with $\sim$ are pseudo-luminosity estimates; other estimates come from $^\mathrm{a}$Galactic Center distance, $^\mathrm{b}$\citep[SNR velocity]{reynoso2006high}, $^\mathrm{c}$\citep[dispersion measure, DM]{ya02017new}, $^\mathrm{d}$\citep[H$\alpha$ bow shock]{de2020psr}.}
    \label{tab:results}
\end{table*}

\begin{figure*}
    \centering
    \includegraphics[width=\linewidth]{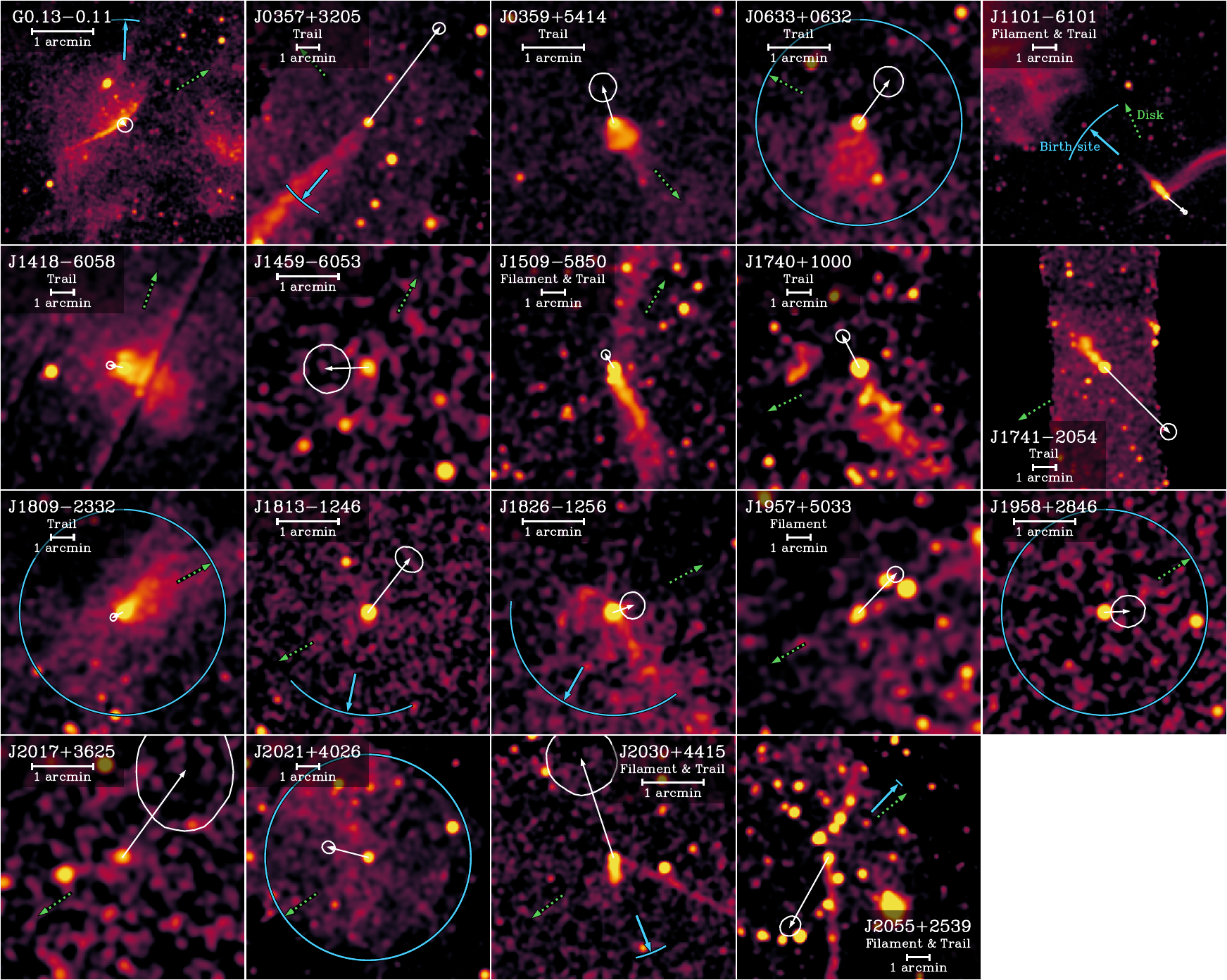}
    \caption{\textit{Chandra} images of the pulsars considered in this work. Our best-fit proper motions are shown as white arrows, scaled to 1 kyr of motion, with the arrow tip enclosed by 68\% confidence intervals. The directions to the Galactic plane and potential birth sites are marked in green and blue respectively. Blue arcs show the angular extent of the birth site. A blue circle indicates that the pulsar is inside the birth site.}
    \label{fig:results}
\end{figure*}

If one assumes zero X-ray background ($f_s = 1$), then $\ln L$ is proportional to the traditional log-Poisson figure of merit statistic \citep[FoM;][]{van2012chandra} applied to an infinitely fine grid, with exposure map correction. Our inclusion of the X-ray background is important because it enables fitting to X-ray-faint sources and pulsars embedded in bright nebulae. The individual $f_s$ would ideally be treated as fit parameters, but for a fit with hundreds of PSs this is infeasible. Instead we take an iterative approach. We perform an initial fit for $f_s$ by maximizing $L$ while setting the fit parameters to zero. Fixing $f_s$ to this best-fit value, we then fit $\bm \delta_f$ and $\dot {\bm \delta_f}$ to the X-ray bright PSs with \textit{Gaia} counterparts. Now that the frames are better aligned, we refit $f_s$ and remeasure the sources' colors $c_X$, and then, using the spectrum-appropriate PSFs, simultaneously fit $\bm \delta_f$, $\bm x_s$ and $\dot {\bm \delta_f}$ to all PSs. Finally, we refit $f_s$, remeasure $c_X$ again, and simultaneously fit $\bm \delta_f$, $\bm x_s$, $\dot {\bm \delta_f}$ and $\bm \mu_s$.

To make this repeated minimization numerically tractable, we pre-compute the first and second derivatives of $P_{s,f}$ and $C_{\mathrm{exp},f,s}$ as a function of the fit parameters. The derivatives of $L$ are linear combinations of these derivatives, which we use to efficiently minimize $L$ with Newton's method and estimate a Gaussian covariance matrix $\Sigma$ from the log likelihood: $(\Sigma^{-1})_{ij} = -\partial_i \partial_j \ln L$. This method of determining $\Sigma$ appropriately propagates $\bm \delta_f$, $\bm x_s$, and $\dot {\bm \delta_f}$ uncertainties to the final pulsar PM uncertainty.

\section{Pulsar Birth Sites}
\label{sec:physics}
Our PM measurements and uncertainties allow us to propagate pulsars back through the Galactic potential, identifying the paths they may have traversed during their spin-down age. The supernova remnant (SNR) birth site may still be visible for pulsars $< 100$ kyr old; otherwise, OB associations are the best bet for identifying birth sites.  SNRs or OB associations that cross the possible pulsar tracks are identified as birth sites. We use the SNR catalog of \cite{ranasinghe2022distances} and the \textit{Gaia}-based OB association catalog of \citep{chemel2022search}, both of which give distances and radii.

Estimates of the current velocities of the pulsars, OB associations, and SNRs are needed to propagate the objects through the Galactic potential. The OB association catalog gives PM measurements, and we extract the unmeasured components assuming that the objects follow the local Galactic rotation. The pulsar's radial velocity is treated as a  double Gaussian corresponding to the double Maxwellian velocity distribution found by \cite{igoshev2020observed}. We center the radial velocity distribution so that the mean is the line-of-sight component of the local Galactic circular velocity. Orbits are integrated using the \texttt{galpy} \texttt{Python} package \citep{bovy2015galpy}.

To quantify the significance to birth site identifications, for each time $t$ and each three-dimensional birth site track $\bm x(t)$ we calculate the ``odds statistic'' $\mathcal O(t) = P_\mathrm{psr}(\bm x(t))P_\mathrm{psr}(t) / \rho_\mathrm{site}(\bm x(t))$. $P_\mathrm{psr}(\bm x)$ is the normalized probability density function that the pulsar was born at position $\bm x$, determined from the cloud of simulated tracks. $P_\mathrm{psr}(t)$ is the prior probability that the pulsar was born at time $t$ ago. $\rho_\mathrm{site}(\bm x)$ is the local density of birth sites, which we estimate by convolving the catalogs' birth site distributions with a wide Gaussian. Large odds indicate a likely association (high pulsar probability $P_\mathrm{psr}(x)P_\mathrm{psr}(t)$, and/or a low false positive rate $\rho_\mathrm{site}(\bm x)$). The $\mathcal O$ statistic is then converted to a $p$-value, defined as the probability that sites distributed randomly over $\rho_\mathrm{site}$ could produce a value of $\mathcal O$ even greater than $\max_t \mathcal O(t)$.\footnote{This is a frequentist approach. A Bayesian analysis would give similar results because $\mathcal{O}$ is the Bayes factor using a uniform prior on the pulsar's position and $P_\mathrm{psr}(t)$ as a prior on time.} We calculate $p$ using the probability density of $\mathcal{O}$ extracted from the odds map. Small $p$-values correspond to better detections.

Since irregular explosions and non-uniform surrounding can center the present SNR limb away from the explosion site, we convolve $P_\mathrm{psr}(\bm x)$ with a Gaussian of standard deviation equal to a tenth of the SNR radius, to account for possible offsets. For OB associations, we use a Gaussian with standard deviation equal to the radius containing 68\% of the cluster members. Our heuristic prior on the time of birth is piecewise. When $t \leq $ the spin down age $\tau_c= P/(2\dot P)$, we set $P_\mathrm{psr}(t) \propto t/\tau_c$ to prefer associations that give ages closer to the spin-down age. When $\tau_c \leq t < 2\tau_c$ we maintain a flat prior $P_\mathrm{psr}(t)=1$ to allow possible birth earlier than the spin-down age. We omit associations with $t>2\tau_c$ by setting $P_\mathrm{psr}(t)=0$.

When an association is available, the time of intersection presents an estimate of the kinematic age $\tau_k$. For constant magnetic dipole spin-down, the pulsar's initial period is $P_0 = P\sqrt{1 - \tau_k / \tau_c}$ where $P$ is the current spin period \citep{goldreich1969pulsar}. When no association is present, we can still estimate the kinematic age by assuming that the pulsar was born from high mass progenitors near the Galactic disk (although some are certainly born from O star runaways and other high $|z|$ sources). Calculating when the pulsar tracks intersect the disk, allowing a $\sim$50 pc exponential scale height progenitor distribution, gives a disk $\tau_k$ and initial period; we only report a disk $\tau_k$ when $>50\%$ of the tracks cross the disk.

Most of the nineteen pulsars we consider show a clear X-ray trail opposing the PM. For the remaining six, we search for faint trails. This is done by summing the 0.5$-$7 keV \textit{Chandra} counts in a $20''\times 8''$ rectangular aperture opposing the best-fit PM direction, excluding counts within $3''$ of the pulsar. This size was chosen to match the Guitar pulsar's faint X-ray trail (detected to 4.3$\sigma$ with this method). In all cases, an excess above the local background was found. We convert the count rate to a significance assuming Poisson statistics. When a trail is detected, we also present an absorbed flux estimate. To convert from the observed counts to energy, we assume an power law $dN_\gamma/dE_\gamma \propto E_\gamma^{-\Gamma}$ with photon index $\Gamma=2$, absorbed with the \texttt{xspec} \texttt{tbabs} model under column density $n_H = 5\times 10^{21}$ cm$^{-2}$. Low significance excesses are reported as 90\% upper limits to the flux.

\section{Results}
\label{sec:results}

\begin{figure*}
    \centering
    \includegraphics[width=0.8\linewidth]{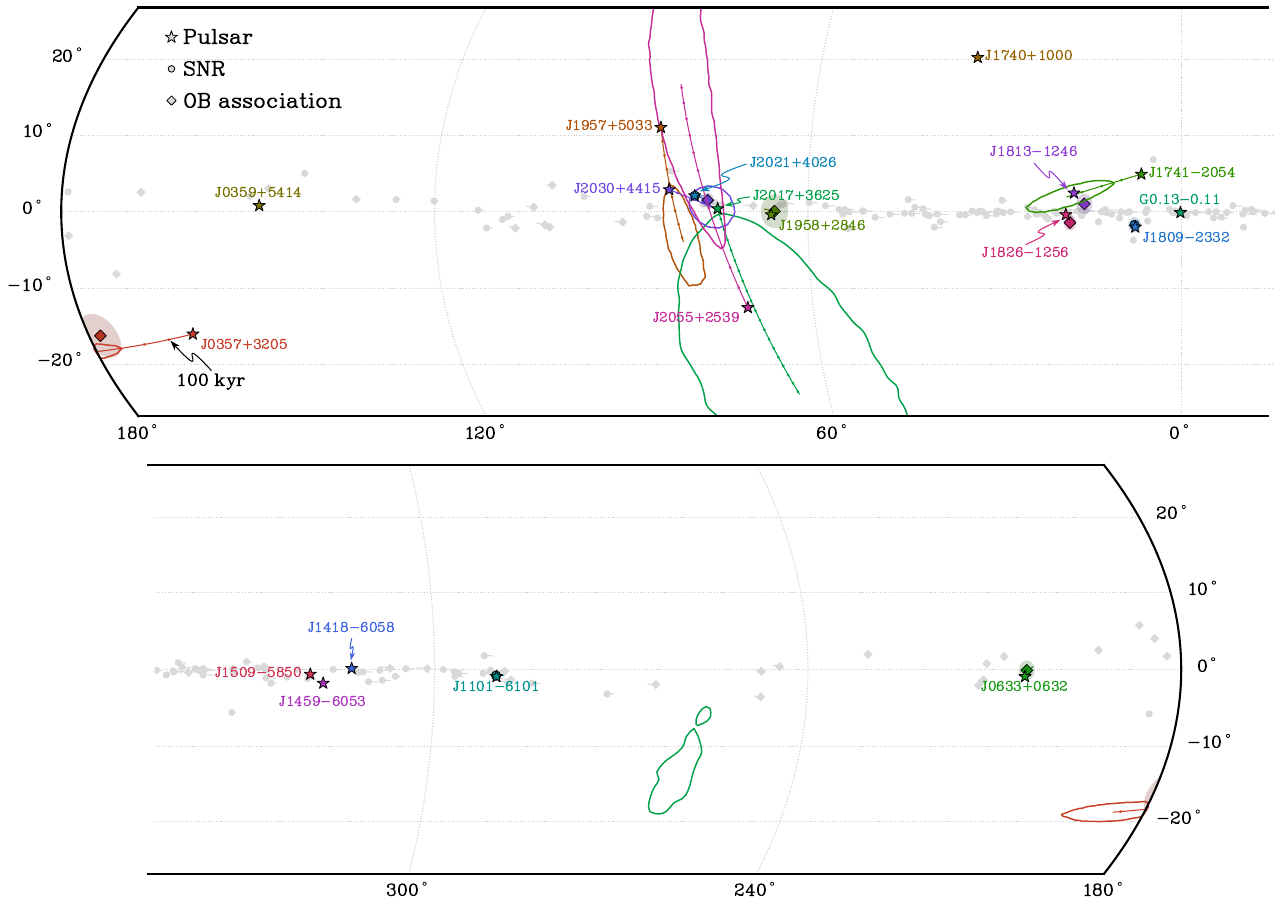}
    \caption{Locations and tracks of the pulsars considered in this work, in Galactic coordinates. Tracks are determined by propagating the observed proper motion through the Galactic potential and terminating at the spin-down age. 68\% confidence intervals represent combined proper motion, radial velocity, and 30\% distance uncertainties. SNRs and OB associations within 3 kpc of the Sun are also shown. When a pulsar track intersects with an SNR or OB track, the association is marked with the pulsar's color and the size is shown.}
    \label{fig:glgb}
\end{figure*}

Our PM measurements are presented in Table \ref{tab:results}. Components $\mu_x$ and $\mu_y$ are measured in the directions of increasing right ascension and declination. We measure pulsar positions to an accuracy of tens of milliarcseconds on the \textit{Gaia} reference frame, and achieve PM uncertainties of a few mas yr$^{-1}$. Most pulsars produced significant PM. We also give a plane-of-sky velocity using a reliable distance measurement when available and otherwise the $\gamma$-ray pseudo-luminosity distance $d = (10^{33}\ \mathrm{erg}\ \mathrm{s}^{-1} \dot E)^{1/4} / (4\pi F_\mathrm{\gamma})^{1/2}$, where $F_\gamma$ is the \textit{Fermi} Large Area Telescope $\gamma$-ray flux reported in the 4FGL-DR4 catalog \citep{abdollahi2020fermi}. As PSR J1101$-$6101 is not $\gamma$-ray detected (the $\gamma$-ray emission is likely associated with a nearby flat spectrum radio quasar), we use the distance of the associated SNR to calculate a velocity. PM significances are also reported, based on the fact that $\mu_x^2 / \sigma_{\mu_x}^2 + \mu_y^2 / \sigma_{\mu_y}^2$ is $\chi^2$-distributed with two degrees of freedom.

Fig.~\ref{fig:results} shows \textit{Chandra} images of each pulsar, with the arrow showing the pulsar shift over 1 kyr, and 68\% confidence intervals surrounding the arrow tip. The direction to the Galactic disk is also indicated, and the direction to a potential birth site if available.

In Fig.~\ref{fig:glgb} we present an overview of the pulsar motions relative to the (slowly changing) direction from the Earth to the Galactic center, including the Solar peculiar velocity. The tracks for pulsars that have shifted by $>1^\circ$ show apparent motion across the sky and culminate in 68\% positional confidence intervals at the spin-down age. Note that the contour of J2017+3625 wraps under the map, so that it has substantial support at both $\ell \sim 60^\circ$ and $\ell \sim 260^\circ$. Since the spin-down age is not always an accurate estimate of true age, we mark $10^5$ year intervals along the tracks where appropriate. Motions and uncertainties are most prominent for the nearby pulsars. 

The SNRs and OB associations discussed in \S\ref{sec:physics} within 3 kpc of the Sun are shown in Fig.~\ref{fig:glgb}, with tracks reflecting 1 Myr of their apparent motion. When a birth site is likely affiliated with a pulsar, it is marked with the pulsar's color and presented in Table \ref{tab:assoc}. G0.13$-$0.11's track angle and length are unknown due to the object's poorly constrained PM and unmeasured pulsations, so we do not search for birth sites.

Birth site identification requires knowledge of the pulsar's distance, which is usually uncertain. We assume 30\% distance uncertainties as a fiducial value, but we also search for birth sites with wider, 2 kpc distance uncertainties. As PSR J1101$-$6101 has no $\gamma$-ray distance, we assume very wide 6.3 kpc uncertainties. When the wider search presents new results, the associations are reported in Table \ref{tab:assoc} in parentheses. When an association is detected with $p$-value $<20\%$, we estimate the kinematic age from the the time of intersection with the birth site track. Upper and lower bounds are the range in which the $p$-value doubles.

For pulsars traveling away from the Galactic plane, the crossing time also gives a kinematic age estimate in Table \ref{tab:assoc}.  Of course, as displayed in Figure \ref{fig:glgb} some pulsars such as J0357 are headed {\it toward} the plane. These may have arisen on one of the higher $|b|$ OB associations, or from runaway O stars released from binaries via the Blaaw mechanism and exploding at high $|b|$. Their pulsar progeny can be driven back toward the disk via birth kicks. Finally, Table \ref{tab:assoc} also presents the significance and flux upper limit of our search for X-ray trails opposing the PM. We discovered three X-ray trails to $>3\sigma$ significance. Fluxes for low confidence detections are reported as upper limits.

\begin{figure}
    \centering
    \includegraphics[width=\linewidth]{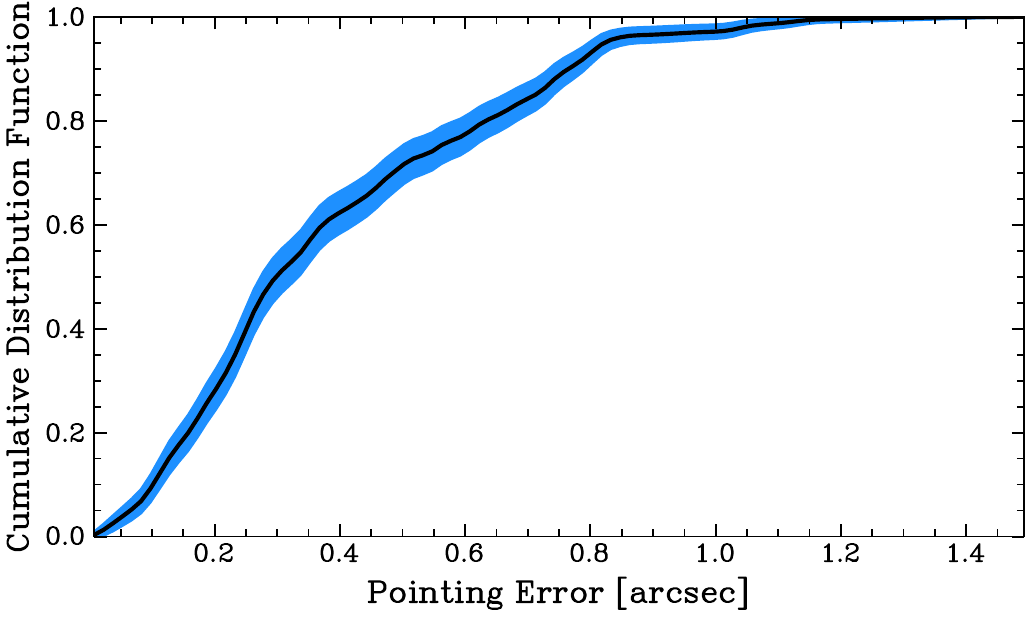}
    \caption{Distribution of \textit{Chandra} frame-frame pointing offsets measured by our method. The blue band indicates uncertainties.}
    \label{fig:errors}
\end{figure}

A byproduct of our analysis is milliarcsecond-precise estimates of the pointing error of each frame. Fig.~\ref{fig:errors} shows the pointing error distribution of our observations. The error band includes both the uncertainty of each pointing propagated from our fit results, and the uncertainty on the true distribution stemming from the finite number of observations. Results show a median offset of $0.30''$ and 90\% upper limit of $0.77''$.

\begin{table*}
    \footnotesize
    \centering
    \begin{tabular}{llccccccccc} \hline \hline 
        Name & Birth site & $p$ & $d_\mathrm{psr}/d_\mathrm{site}$ & $P$ & $\tau_c$ & $\tau_k$ / $P_k$ & $\tau_\mathrm{disk}$ / $P_\mathrm{disk}$ & Trail & Trail \\
        & & [\%] & [kpc]/[kpc] & [ms] & [kyr] & [kyr] / [ms] & [kyr] / [ms] & flux & $Z$ \\ \hline 
        J0357+3205 & \textbf{OB G178.1$-$16.2} & 1.8 & 0.6 / 0.6 & 444 & 540 & $330^{+90}_{-40}$ / $280^{+20}_{-60}$ &&  &  \\ 
        J0359+5414 &  &  &  & 79 & 75 && $200^{+100}_{-100}$ / $50^{+30}_{-20}$ &  &  \\ 
        J0633+0632 & OB G204.8$-$0.1 & 12.9 & 1.0 / 1.2 & 297 & 59 & $60^{+60}_{-20}$ / $0^{+200}$ &&  &  \\ 
        J1101$-$6101 & \textbf{(SNR G290.1$-$0.8)} & (0.03) & (6.3 / 6.3) & 63 & 116 & ($13^{+4}_{-1}$ / $59^{+0}_{-1}$) & $120^{+50}_{-60}$ / $30^{+20}_{-10}$ &  &  \\ 
        J1418$-$6058 &  &  &  & 111 & 10 &&&  &  \\ 
        J1459$-$6053 &  &  &  & 103 & 65 && $800^{+900}_{-600}$ / $70^{+30}_{-30}$ & $<2.3$ & 0.3 \\ 
        J1509$-$5850 &  &  &  & 89 & 154 &&&  &  \\ 
        J1740+1000 &  &  &  & 154 &  &&&  &  \\ 
        J1741$-$2054 &  &  &  & 414 & 386 && $900^{+1000}_{-700}$ / $280^{+130}_{-60}$ &  &  \\ 
        J1809$-$2332 & \textbf{(SNR G7.5$-$1.7)} & (0.2) & (0.6 / 1.7) & 147 & 68 & ($60^{+0}_{-20}$ / $51^{+35}_{-2}$) & $1200^{+1100}_{-900}$ / $100^{+40}_{-40}$ &  &  \\ 
        J1813$-$1246 & OB G15.6+1.0 & 12.7 & 1.6 / 1.6 & 48 & 43 & $80^{+10}_{-40}$ / $0^{+10}$ & $300^{+200}_{-200}$ / $30^{+20}_{-10}$ & 5(4) & 3.3 \\ 
        J1826$-$1256 & (OB G17.9$-$1.4) & (13.8) & (1.1 / 2.0) & 110 & 14 & ($25^{+4}_{-12}$ / $0^{+32}$) &&  &  \\ 
        J1957+5033 &  &  &  & 375 & 838 && $600^{+200}_{-300}$ / $230^{+60}_{-60}$ & $<1.3$ & 1.6 \\ 
        J1958+2846 & (OB G65.4+0.1) & (18.1) & (1.2 / 2.4) & 290 & 22 & ($36^{+3}_{-10}$ / $-$) && 3(2) & 3.0 \\ 
        J2017+3625 &  &  &  & 167 & 1940 && $200^{+300}_{-200}$ / $160^{+8}_{-4}$ & $<1.5$ & 1.2 \\ 
        J2021+4026 & OB G83.1+0.2 & 10.7 & 0.3 / 0.4 & 265 & 77 & $80^{+80}_{-30}$ / $0^{+100}$ && 5(4) & 3.3 \\ 
         & \textbf{(SNR G78.2+2.1)} & (0.2) & (0.3 / 2.1) &  &  & ($6^{+6}_{-5}$ / $279^{+8}_{-10}$) &  &  &  \\ 
        J2030+4415 & \textbf{OB G76.1+1.5} & 5.3 & 0.8 / 1.0 & 227 & 555 & $420^{+100}_{-50}$ / $110^{+20}_{-50}$ & $800^{+800}_{-600}$ / $150^{+70}_{-30}$ &  &  \\ 
        J2055+2539 & OB G74.6+4.6 & 15.1 & 0.6 / 1.1 & 320 & 1230 & $1200^{+200}_{-300}$ / $50^{+120}_{-50}$ & $500^{+300}_{-300}$ / $250^{+40}_{-20}$ &  &  \\ 
        B2224+65 &  &  &  & 683 & 1120 &&&  &  \\ 
        \hline \hline 
    \end{tabular}
    \caption{Physical quantities derived from the PM measurement. From left to right, possible SNR / OB association birth sites, the $p$-value of association (small numbers represent more confident assignments), the pulsar / birth site distance, the current spin period, the spin down age, estimates of the pulsar age/initial period from intersection with the birth site, the same for disk crossing, absorbed X-ray flux of faint trails in 10$^{-15}$ erg cm$^{-2}$ s$^{-1}$, and trail significance. We do not give trail fluxes and significances for bright trails previously noted in the literature as a more detailed spectral analysis is required to give accurate fluxes. Birth sites with $p<10\%$ are the most significant (in bold).}
    \label{tab:assoc}
\end{table*}

\subsection{Individual Sources}

{\bf G0.13$-$0.11} has an X-ray point source, proposed as a pulsar, at the cusp of an X-ray filament with that name, which is located at the tip of the radio arc in the Galactic center. We set a 90\% upper limit of 160 km s$^{-1}$, assuming a distance of 8 kpc. If this compact object is a filament pulsar, it is very likely the slowest. It is possible that large gas density compresses the bow shock enough to spark a filament due to its unique location in the Galactic Center (a number density of $n=3.5$ cm$^{-3}$ is required to reach gate fraction of 30 at the central value of $v_\perp=90$ km s$^{-1}$ and $B=3\ \mu$G, using Eq.~\ref{eqn:gate} discussed further in \S\ref{sec:systematics}). But as pulsations have not yet been detected, G0.13$-$0.11 may also represent a different class of object.

An X-ray PM has already been reported for the spectacular trail of PSR \textbf{J0357$+$3205} to be $(\mu_x, \mu_y)=(-117\pm 20, 115 \pm 20)$ mas yr$^{-1}$ \citep{de2013fast}. Our results are consistent to $\sim 1\sigma$,
with uncertainty reduced by $4\times$ using a new 14 ks observation $\sim 15$ years after the initial epoch. The PM is oriented towards the disk, so it is encouraging that our analysis selects the high latitude OB G178.1$-$16.2 as a birth site, with a plausible $\sim 330$\ yr kinematic age. This implies that it was born as a relatively slow rotator with $P_0 >200$ ms.

PSR \textbf{J0359+5414}, with the ``Mushroom'' PWN exhibits one of the smallest PMs in our catalog. However, at a large distance of 3.9 kpc its velocity of $250 \pm 80$ km s$^{-1}$ is still above the median velocity of $\sim 220$ km s$^{-1}$ \citep{igoshev2020observed}. Given Mushroom's low Galactic latitude of $b = 0.88^\circ$ it should lie at height $z \sim 60$ pc above the Galactic disk and may lie in the cold neutral medium (CNM) ISM phase, which has scale height $\sim 50$ pc \citep{dickey2022gaskap}. The speed of sound is lower in the cold, dense CNM than in other ISM phases, which may help formation of its impressive bow shock.

PSR \textbf{J0633+0632}'s low $\gamma$-ray distance gives a small velocity, but with Galactic latitude of $b=-0.93^\circ$, it is within $\lesssim 50$ pc of the disk. This pulsar shows the poorest alignment between our PM and the trail axis, but some scatter is to be expected in a sample of six clear X-ray trails. Our result is still within 1$\sigma$ of the cone-shaped trail's edges. The pulsar lies inside the possible OB association, and its very rough $\tau_k$ makes the association probable if it was born in the association outskirts and has been kicked toward its center. 

PSR \textbf{J1101$-$6101} is associated with the ``Lighthouse'' X-ray filament. X-ray filament pulsars are expected to have large velocity and be capable of accelerating particles to synchrotron X-ray-emitting energies \citep{bandiera2008on,dinsmore2024catalog}. PSR J1101 was once claimed to be the fastest pulsar, based on the age of the associated SNR MSH 11$-$61A/G290.1$-$0.8 \citep[visible in Fig.~\ref{fig:results}]{tomsick2012is}. Our analysis finds $p = 3\times 10^{-4}$, confirming this this SNR as the likely the birth site, and implies a young $\sim$13 kyr age. Our velocity of $990\pm 40$ km s$^{-1}$ based on the SNR distance is indeed high, but not record-breaking. Our direction is consistent with that indicated by the X-ray trail, and the low $\tau_k$ indicates that it was born near its present spin period. 

PSR \textbf{J1418$-$6058} in the ``Rabbit'' PWN has a significant $\mu=$16\,mas yr$^{-1}$, yielding a sub-median velocity. This pulsar's Galactic latitude is so low at $b=0.13^\circ$ that its $\gamma$-ray distance puts it 3 pc above the disk, plausibly residing in the CNM. The pulsar's bright trail is unusually wide and diffuse. Together with the low PM, this suggests we may be viewing at small inclination, with the pulsar having substantial radial velocity.

The PM of PSR \textbf{J1459$-$6053} is tentatively measured at 2.2$\sigma$ with 20\,ks of new data providing a baseline of 13 years. This signal would have gone undetected without the use of \textit{Gaia} stars, as Fig.~\ref{fig:field} shows that frame uncertainties are tripled if only X-ray point sources with $\gtrsim 10$ total counts are used. The velocity of $140 \pm 50$ km s$^{-1}$ is modest and we derive only upper limits on extended emission.

PSRs \textbf{J1509$-$5850} is a two-tailed PWN, with one structure representing a trail and the other a filament. Our velocity measurement conclusively demonstrates that the northern structure is the filament. This assignment was first suggested by \cite{klingler2016chandra} because the apex of the southern structure resembles a compact PWN. The velocity itself is large, typical of X-ray filament pulsars.

PSR \textbf{J1740$+$1000} is at high Galactic latitude ($b=20.3^\circ$) and its PM is parallel to the disk. Unlike J0357, we have not identified a likely OB/SNR birth site, but it could have arisen from an O star runaway. Both trail \citep{kargaltsev2008xray} and filament \citep{dinsmore2024catalog, gagnon2025nature} interpretations have been proposed for its PWN. The presence of the TeV source 1LHAASO J1740+0948u $\sim 13'$ behind the pulsar \citep{brunelli2025investigating} supports the trail picture and our PM measurement confirms this interpretation. Based on our $46\pm 5$ mas yr$^{-1}$ PM and \cite{cao2024first}'s $4.2'$ 95\% confidence uncertainty on the TeV source's position, the two were coincident $\sim 17 \pm 4$ kyr ago. If the pulsar is responsible for the TeV emitting electrons, these can be the cooled relic of a population emitted at the TeV site when the pulsar was young, or freshly injected electrons ducted along the pulsar trail \citep{brunelli2025investigating,gagnon2025nature}. Both seem viable, although with $\tau_k\sim 0.1 \tau_c$ the pulsar luminosity at birth would not be much higher than at present (unless initial spin-down was very rapid). Also the inverse Compton cooling time is $4/E_{100\,\mathrm{TeV}}$\,kyr, so while the kinematic age can accommodate the $\sim 30$\,TeV LHAASO peak, maintaining electrons to provide the highest energy TeV photons may require replenishment via advection along the trail.

The PM of \textbf{J1741$-$2054} was previously reported to be $(-63\pm 12, -89\pm 9)$ mas yr$^{-1}$ by \cite{auchettl2015xray}. Our result is consistent with this result to 1.5$\sigma$ in RA and 0.4$\sigma$ in DEC, and aligns well with the trail axis. Our uncertainties are smaller by 30\%; this improvement is due to our use of \textit{Gaia} stars, as we collected no new data. At the DM-estimated 0.27\,kpc, the velocity is below average at 150 km s$^{-1}$.

Likewise, a PM has been reported for \textbf{J1809$-$2332} as $(12\pm 8, -24 \pm 6)$ mas yr$^{-1}$ \citep{van2012chandra}. With an additional 15 ks observation made 13 years after the previous epoch, we match the $\mu_x$ measurement and reduce uncertainties by a factor of 2.2. However our $\mu_y$ measurements are in 2.7$\sigma$ tension. Our PA is $118\pm 13^\circ$, while that of \citep{van2012chandra} is $153\pm 16^\circ$. Both are consistent with opposition to the X-ray trail, and ours is closer to opposition with SNR G7.5$-$1.7 at PA$=(132+180)^\circ$, previously proposed as the birth site \cite{van2012chandra}. The association is highly likely at $p=0.2\%$, if the PSR DM distance estimate is $\sim 3\times$ too low. This would also boost the pulsar transverse velocity to $\sim 130$ km s$^{-1}$ from an unusually low 40 km s$^{-1}$.

PSR \textbf{J1813$-$1256} exhibits a large PM of $33\pm 4$ mas yr$^{-1}$, which is quite significant despite the short 18 ks new exposure. This pulsar's frame alignment benefits substantially from the use of \textit{Gaia} stars (Fig.~\ref{fig:field}). At the $\gamma$-ray distance of 1.6 kpc, our PM corresponds to a typical velocity of $260 \pm 30$ km s$^{-1}$ and we have a low significance ($p=13\%$) birth site, but with reasonable kinematic age if the pulsar was born as a fast rotator. We discover a trail at 3.3$\sigma$ significance.

The PM of PSR \textbf{J1826$-$1256} is particularly important for interpreting its unusual PWN. Under the assumption that the two thin structures extending from the pulsar are jets, \cite{karpova2019xray} highlighted two possible explanations. If the PM points north then the bend in the eastern jet could arise from ISM ram pressure, and western jet is straighter because it trails the pulsar as first suggested by \cite{roberts2009pulsar}. Another configuration suggested by \cite{kargaltsev2017pulsar} is akin to that of PSR B0633+17 ``Geminga''--- with both jets equally swept back by ISM ram pressure, but, with the structure inclined to the line of sight, the eastern jet appears more bent due to projection effects. Our PM measurement is perpendicular to both jets, and inconsistent with PA=0 to $>2\sigma$. It therefore supports the second explanation. We again have a low-confidence birth site, plausible if the pseudo-luminosity distance estimate is low and the pulsar was born as a fast rotator.

Our 10$\sigma$ significant PSR \textbf{J1957+5033}'s PM is misaligned with the X-ray structure, demonstrating for the first time that the associated X-ray structure is a filament. The filament's width is close to \textit{Chandra}'s resolution, like that associated with J2030+4415. An XMM exposure of J1957 (observation ID 0844930101) reveals a length of $\sim 14.5'$, also comparable to the J2030 filament. A primary difference is J1957's high Galactic latitude of 11$^\circ$---the largest of all known filaments. The presence of a filament is then puzzling as the pulsar is not likely in a dense ISM phase unless $d\lesssim$300\,pc. While the high-resolution dust map of \cite{edenhofer2024parsec} detects no dense dust clouds near the pulsar pseudo-luminosity distance of 1.2 kpc, a smaller dense clump might be present, or the pulsar might lie closer, embedded in observed dust and gas at $\sim 400-600$\,pc. Given the large pulsar $\tau_c$ it is unsurprising that we detect no birth site, but the kinematic age for birth in the Galactic plane is reasonable.

The transverse velocity of PSR \textbf{J1958+2846} is quite low. We do find a nominal origin in a superimposed OB association, but the distance estimates disagree and the kinematic age exceeds the pulsar spin-down age, preventing us from estimating $P_k$. With our lowest association significance ($p=18$\%) this may well be a false positive. We report a $>3\sigma$ X-ray trail detection in Table \ref{tab:assoc}; at $\sim 7$\, pc from the plane, ISM confinement and a PWN trail are not unexpected.

Due to its close distance and large spin-down age of 1.9 Myr, PSR \textbf{2017+3625}'s track uncertainty covers a lot of sky. Nevertheless, no probable affiliation is seen, unsurprising given the large $\tau_c$. The small $\tau_\mathrm{disk}$ indicates either a birth site at large negative $b$ or initial period very close to the present value.

PSR \textbf{J2021+4026} exhibits a large and well-measured PM of $\mu = 53 \pm 5$ mas yr$^{-1}$, but the large $\gamma$-ray flux gives a low pseudo-luminosity distance and, accordingly, a low transverse velocity. At the $\gamma$-ray distance of 0.3\,kpc, we do find a plausible OB association, with a reasonable $\tau_k$. However $\gamma$-Cygni \citep[SNR G78.2+2.1, suggested by ][]{hui2015detailed} is also plausible as a birth site if the distance is greatly underestimated. The larger SNR distance would give the pulsar a more typical transverse velocity of $530\pm 50$ km s$^{-1}$ instead of $84\pm 9$ km s$^{-1}$, and a slow initial spin. We saw a faint X-ray trail opposing the PM to  3.3$\sigma$ significance.

\textbf{J2030+4415}'s X-ray PM was estimated to be $(15\pm 11, 84\pm 12)$\,mas yr$^{-1}$ by analyzing the same X-ray data with the FoM technique \citep{de2020psr}. Our measurement is along a similar axis, and consistent with the direction indicated by the X-ray trail. Uncertainties are comparable because this pulsar's field contains many X-ray bright PSs (Fig.~\ref{fig:field}), and the use of \textit{Gaia} stars offers only marginal improvement. However, our PM magnitude is 2.2$\sigma$ smaller. This lower PM is persistent across different samples of field PSs, and is found even when we exclude X-ray faint stars from the fit. The PM opposes the direction to OB G76.1+1.5 at moderate confidence, but the alignment vanishes if one allows a 1$\sigma$ fluctuation to align the PM with the X-ray trail.

Like J1509$-$5850, PSR \textbf{J2055$-$3625} is a two tailed PWN where one structure likely represents an X-ray trail and the other an X-ray filament. Previously, \cite{marelli2019two} suggested that the northern structure represented a trail because the southern structure's properties were inconsistent with trail models, whereas \cite{dinsmore2024catalog} suggested that the spectrum of the northern structure (with stars removed) was more consistent with a filament. Our PM measurement demonstrates that the assignment of \cite{marelli2019two} was correct. J2055's affiliation with OB G74.6+4.6 is of low confidence but the inferred $\tau_k$ is in good agreement with the pulsar spin-down age.

\section{Review of systematic Uncertainties}
\label{sec:systematics}

To test the robustness of our approach to systematic errors, we re-measured the PM of several pulsars with previous \textit{Chandra} PM measurements as already noted and obtained similar results (with smaller or better uncertainties). We also measured the X-ray PM of the radio-loud PSR B2224+65, associated with the Guitar nebula and X-ray filament. The pulsar's large, known PM and plentiful X-ray data makes it a good probe of systematic error. The resulting X-ray PM of $(143\pm 2,126\pm 2)$ mas yr$^{-1}$ is 2$\sigma$ separated from the precise VLBI measurement of $(147.2\pm 0.2, 126.5\pm 0.1)$  mas yr$^{-1}$. If $1\sigma$ of the tension is attributed to systematic uncertainty, one gets $\sigma_\mathrm{sys} =  2$ mas yr$^{-1}$.

The rest of this section identifies possible sources of systematic error and estimates their effect.  Altogether, the tests suggest $\sim 2$ mas yr$^{-1}$ of systematic error, with lower values for pulsars with faint nebulosity. Errors of this size are subdominant to statistical uncertainties in all cases but J1101$-$6101 and G0.13$-$0.11, and do not affect our birth sites and dynamical age estimates.

\subsection{PWN Fluctuations}
\label{sec:pwn-fluctuations}
PWN termination shocks, such as the Crab wisps, can be variable and fluctuations near the pulsar might randomly shift the best-fit pulsar position by some amount $\Delta \bm x$ in each observation. For pulsars with few observed epochs, this shift would appear as a linear trend and bias the PM. We can measure these shifts in deep exposures by fitting a single PS to the pulsar, and then re-fitting with a fainter PS added to represent a nebular bright spot. In the worst-case scenario of a two-epoch PM fit, these shifts will appear as a linear trend and bias the PM by $\sim \sigma_\mathrm{sys,neb}=|\Delta \bm x| / \Delta t$. For all but the brightest nebulae we find this quantity to be $<1$ mas yr$^{-1}$, subdominant to statistical errors. The exceptions are J1101$-$6101 and J1809$-$2332, which have $\sigma_\mathrm{sys,neb}\approx 2-3$ mas yr$^{-1}$, and J1509$-$5850 with $\sigma_\mathrm{sys,neb}=6$ mas yr$^{-1}$. However, these nebulae have $N>2$ observational epochs, so the PM bias is decreased by $\sim 1/\sqrt{N-1}$ assuming uncorrelated shifts between different epochs. The largest systematic uncertainty is then $\sigma_\mathrm{sys,neb} = 3$ mas yr$^{-1}$ for J1509, compared to the 4 mas yr$^{-1}$ total statistical uncertainty, and $<2$ mas yr$^{-1}$ for all other pulsars.

\subsection{Field PS Mis-modeling}
Another source of error is imperfect modeling of the field PSs, e.g.~if they possessed undetected PMs, nebulosity, or irregular spectra that caused our model PSFs to mis-fit the data. This would invalidate our frame alignments. X-ray bright PSs are the most likely source of bias as these relatively few sources contribute strongly to the frame registration. We test for sensitivity to the X-ray bright PS selection excluding the five brightest PSs from the fit one by one and re-measuring the pulsar PM. The PMs show an unpatterned scatter with standard deviation between $1/2$ and $1\times$ the statistical uncertainty. $\sim 1 \sigma$ scatter is expected when sub-sampling the data set. This result therefore indicates that bias from bright PS mis-modeling is small. We chose to use the pulsar to constrain frame alignment assuming constant pulsar PM, so pulsar mis-modeling could also cause systematic error in frame alignment. However, \S\ref{sec:pwn-fluctuations} showed that the pulsar's motion is consistent with a constant PM, so this choice should not contribute significant systematic error to the frame alignment.

\subsection{Spacecraft Roll Error}
We have assumed that the reported spacecraft roll angles are perfectly accurate, but the \textit{Chandra} observatory guide reports 25$''$ roll angle uncertainty. If a cluster of bright field PSs is offset from the pulsar, our fit method will compensate for a roll offset by translating the frame such that the bright PSs remain stationary. This would affect the pulsar proper motion. PSR J1809$-$2332 is the most likely source to show this effect because it is 6.5$'$ offset from the 7 Myr-old open cluster Collinder 367, which contributes 21 of our 50 X-ray bright field PSs. We calculate the systematic uncertainty due to roll error for this pulsar by offsetting the X-ray events corresponding to a roll error $\delta \theta$ and re-measuring the pulsar PM. $\delta \theta$ is drawn from a Gaussian distribution with 25$''$ standard deviation. We repeat this process and interpret the PM scatter as the systematic error due to roll offsets. Our result of $\sigma_\mathrm{sys,roll} = 0.9$ mas yr$^{-1}$ is subdominant to the statistical error for J1809 of 3 mas yr$^{-1}$. Since no other field PS distributions show similarly extreme asymmetry, we expect that spacecraft roll does not contribute significant systematic error.

\section{Discussion \& Conclusions}
\label{sec:conclusion}
The new techniques introduced in this paper leverage twenty-five years of \textit{Chandra} pulsar observations and \textit{Gaia} astrometry to deliver accurate pulsar positions and the most precise X-ray PMs yet measured. Our premise is that X-ray faint \textit{Gaia} stars provide more counts to aid in frame registration, as does the pulsar when simultaneously fitted. These improvements are particularly useful in fields with few X-ray bright point sources.

A principal application of these motions is to identify potential birth sites. We have identified three parent SNRs and two OB associations with high confidence, and an additional six OBs at lower confidence. These identifications also give an estimate of the birth spin period (for constant dipole braking).  In two cases (J0357 and J2021) these initial spin periods are over 200\,ms and several more have initial spins larger than 50\,ms, reinforcing the evidence that some pulsars are born slowly rotating. 

The velocity vectors also help in interpreting associated PWN morphology. The vectors indicate ram pressure confined PWN trails are visible for most objects, with two new faint trails discovered in our new data. Confinement seems especially effective for pulsars at low $|z|$, where ISM densities are large. The effects of confinement and sweep-back appear particularly dramatic for J1826, where our measured PM suggests that we are seeing swept back polar jets.  

Precise velocities also illuminate the conditions necessary to produce X-ray filaments. The traditional picture is that the pulsar bow shock traps low energy particles from escaping to the filament, but escape becomes more likely when the particle Larmor radius $r_L$ exceeds the bow shock stand-off distance $r_0$. Defining $\gamma_\mathrm{esc}$ to be the Lorentz factor at which $r_L = r_0$, \citep{dinsmore2024catalog} introduces the ``gate fraction''
\begin{equation}
    \begin{aligned}
    g &= \frac{\gamma_\mathrm{MPD}}{\gamma_\mathrm{esc}} \\
    &= 38 \parens{\frac{v_\perp}{300\ \mathrm{km}\ \mathrm{s}^{-1}}}\parens{\frac{n_\mathrm{ISM}}{0.5\ \mathrm{cm}^{-3}}}^{1/2}\parens{\frac{B}{3\ \mu\mathrm{G}}}^{-1}
    \label{eqn:gate}
    \end{aligned}
\end{equation}
as a metric to predict whether filaments can form. $\gamma_\mathrm{MPD}$ is the maximum Lorentz factor accelerated by the pulsar maximum potential drop, $n_\mathrm{ISM}$ refers to the ISM gas number density, and $B$ is the ISM magnetic field strength. Under this model, the pulsar spawns a filament if $g$ is above some critical value $g_c$, which \cite{dinsmore2024catalog} estimate to be $g_c \sim 35-70$ based on then-known filament velocities. With new velocity measurements for five filaments, we may now improve this bound. We obtain $g=102\pm 24$, $35\pm 11$, $36\pm 9$, $20\pm 6$, and $29\pm 7$ for J1101, J1509, J1957, J2030, and J2055.\footnote{To produce these estimates, we have assumed $B=3\ \mu$G and use the ISM phase fill factors and densities determined by \cite{brownsberger2014survey} as a function of Galactic height, as in \cite{dinsmore2024catalog}. We assume that these filament pulsars lie in one of the neutral ISM phases. The quoted errors propagate uncertainties in PM or the local ISM phase, but not those in distance or $B$.} Uncertainties are appreciable because the distances and local gas density are poorly known; for this reason J2030's low gate fraction was not conclusively constraining in \cite{dinsmore2024catalog}'s $g_c$ estimate. Now that three more filaments also exhibit low gate fraction, $g_c \sim 30$ is a more likely cutoff, making a population of undetected X-ray filaments more likely.

\cite{dinsmore2024catalog} also reported the gate fractions of many pulsars without filaments, to bound $g_c$ from below. Most of these pulsars have $g < 30$, but PSR J1539$-$5626 with $g=53 \pm 34$ and PSR J0248+6021 with $g=33\pm 20$ challenge the gate fraction picture. No filament was detected in 25 ks of \textit{Chandra} data for either pulsar. If either pulsar possesses an undetected filament, it would have to be fainter than any known filament to have evaded detection in \cite{dinsmore2024catalog}. Alternatively, the rarefied hot ionized medium phase has volume filling fraction $>50$\% at the pulsars' Galactic height and could host both pulsars. This would drop their gate fraction to zero. More pulsar velocities are needed to determine whether there is a robust population of high-$g$ pulsars with no filament. If present, such a population would challenge the gate fraction picture. Otherwise, the observed critical gate fraction of $g_c \sim 30$ can be used to constrain the physical mechanisms of filament particle acceleration.

\begin{acknowledgments}
This work was supported in part by NASA grants G05-26045X and G04-25037B administered through the Smithsonian Astrophysical Observatory.
\end{acknowledgments}

\vspace{5mm}
\facilities{CXO}
\software{\texttt{Galpy} \citep{bovy2015galpy}}

\appendix

\section{Observations}
\label{app:obs}

The \textit{Chandra} observations used in this analysis are listed in Table \ref{tab:obs} and are available for download in~\dataset[DOI: 10.25574]{https://doi.org/10.25574/cdc.524}. New observations collected for this study are marked with an asterisk.

\begin{table*}
    \centering
    \begin{tabular}{lcc|lcc|lcc|lcc}
        \hline \hline
         Obs No. & $\Delta t$ [yr] & Exp.~[ks] & Obs No. & $\Delta t$ [yr] & Exp.~[ks] & Obs No. & $\Delta t$ [yr] & Exp.~[ks] & Obs No. & $\Delta t$ [yr] & Exp.~[ks] \\ \hline 
         \multicolumn{3}{c}{\textit{PSR G0.13$-$0.11}} & 24366 & 22.25 & 15 & 29836* & 22.93 & 14 & \multicolumn{3}{c}{\textit{PSR J2017+3625}} \\ 
         945 & 0.00 & 49 & 27463 & 22.26 & 15 & 30729* & 23.07 & 16 & 14699 & 0.00 & 10 \\ 
         13508 & 11.03 & 32 & \multicolumn{3}{c}{\textit{PSR J0357+3205}} & \multicolumn{3}{c}{\textit{PSR J1459$-$6053}} & 29842* & 12.75 & 30 \\ 
         12949 & 11.04 & 59 & 12008 & 0.00 & 30 & 13288 & 0.00 & 10 & \multicolumn{3}{c}{\textit{PSR J2021+4026}} \\ 
         13438 & 11.06 & 67 & 11239 & 0.00 & 47 & 29837* & 13.07 & 10 & 11235 & 0.00 & 56 \\ 
         14897 & 13.08 & 51 & 14207 & 2.17 & 29 & 31956* & 13.07 & 10 & 29843* & 14.41 & 15 \\ 
         17236 & 14.80 & 80 & 14208 & 4.02 & 30 & \multicolumn{3}{c}{\textit{PSR J1509$-$5850}} & \multicolumn{3}{c}{\textit{PSR J2030+4415}} \\ 
         17239 & 15.12 & 80 & 29834* & 15.17 & 14 & 3513 & 0.00 & 40 & 14827 & 0.00 & 25 \\ 
         17237 & 15.86 & 21 & \multicolumn{3}{c}{\textit{PSR J0359+5414}} & 14523 & 10.34 & 95 & 20298 & 4.99 & 45 \\ 
         18852 & 15.86 & 53 & 4657 & 0.00 & 67 & 14524 & 10.60 & 95 & 22171 & 4.99 & 40 \\ 
         17240 & 16.05 & 76 & 14688 & 8.34 & 27 & 14525 & 11.25 & 93 & 22172 & 5.00 & 45 \\ 
         17238 & 17.03 & 66 & 15585 & 8.35 & 22 & 14526 & 11.62 & 95 & 22173 & 5.00 & 23 \\ 
         20118 & 17.04 & 14 & 15586 & 8.35 & 20 & \multicolumn{3}{c}{\textit{PSR J1740+1000}} & 23536 & 6.83 & 25 \\ 
         17241 & 17.24 & 25 & 14689 & 8.41 & 68 & 1989 & 0.00 & 5 & 24954 & 6.84 & 20 \\ 
         20807 & 17.24 & 28 & 14690 & 8.48 & 63 & 11250 & 8.82 & 65 & 24236 & 7.57 & 29 \\ 
         20808 & 17.25 & 27 & 15548 & 8.70 & 67 & \multicolumn{3}{c}{\textit{PSR J1741$-$2054}} & \multicolumn{3}{c}{\textit{PSR J2055+2539}} \\ 
         24362 & 21.57 & 14 & 15549 & 8.75 & 69 & 11251 & 0.00 & 50 & 16957 & 0.00 & 98 \\ 
         26294 & 21.58 & 15 & 15550 & 8.99 & 64 & 14695 & 2.72 & 59 & 16958 & 2.01 & 30 \\ 
         24367 & 21.58 & 18 & \multicolumn{3}{c}{\textit{PSR J0633+0632}} & 14696 & 2.75 & 56 & 29845* & 10.13 & 28 \\ 
         26295 & 21.58 & 12 & 11123 & 0.00 & 20 & 15542 & 2.86 & 29 & \multicolumn{3}{c}{\textit{PSR B2224+65}} \\ 
         24373 & 21.64 & 28 & 19165 & 8.00 & 15 & 15638 & 2.87 & 30 & 755 & 0.00 & 50 \\ 
         24358 & 21.67 & 19 & 20876 & 8.00 & 15 & 15543 & 2.99 & 59 & 6691 & 5.85 & 10 \\ 
         26353 & 21.68 & 10 & 29835* & 15.32 & 15 & 15544 & 3.14 & 57 & 7400 & 5.96 & 37 \\ 
         23641 & 21.75 & 28 & \multicolumn{3}{c}{\textit{PSR J1101$-$6101}} & \multicolumn{3}{c}{\textit{PSR J1809$-$2332}} & 14467 & 11.77 & 15 \\ 
         24365 & 21.76 & 30 & 12420 & 0.00 & 5 & 739 & 0.00 & 10 & 14353 & 11.77 & 35 \\ 
         24370 & 21.78 & 30 & 13787 & 1.10 & 50 & 12546 & 10.95 & 30 & 13771 & 11.78 & 50 \\ 
         26302 & 21.93 & 19 & 16007 & 2.98 & 118 & 29838* & 24.54 & 15 & 24437 & 20.33 & 25 \\ 
         24360 & 21.95 & 15 & 16517 & 3.00 & 53 & \multicolumn{3}{c}{\textit{PSR J1813$-$1246}} & 24434 & 20.39 & 30 \\ 
         26438 & 21.95 & 15 & 16518 & 3.06 & 10 & 14399 & 0.00 & 51 & 24435 & 20.40 & 15 \\ 
         24363 & 21.97 & 28 & 17422 & 3.07 & 50 & 29839* & 11.98 & 18 & 24992 & 20.40 & 15 \\ 
         24369 & 21.99 & 16 & 17421 & 3.07 & 20 & \multicolumn{3}{c}{\textit{PSR J1826$-$1256}} & 24436 & 20.45 & 25 \\ 
         26448 & 22.00 & 17 & 28520 & 12.84 & 58 & 3851 & 0.00 & 15 & 24433 & 20.50 & 26 \\ 
         24368 & 22.03 & 28 & 28519 & 12.85 & 45 & 7641 & 4.44 & 75 & 24431 & 20.50 & 26 \\ 
         24372 & 22.07 & 33 & 28352 & 13.11 & 10 & 29840 & 22.42 & 30 & 24429 & 20.51 & 25 \\ 
         24364 & 22.08 & 11 & 30570 & 13.12 & 12 & \multicolumn{3}{c}{\textit{PSR J1957+5033}} & 24428 & 20.70 & 30 \\ 
         24375 & 22.08 & 28 & 30571 & 13.12 & 10 & 14828 & 0.00 & 25 & 24427 & 20.76 & 25 \\ 
         24750 & 22.15 & 11 & 30572 & 13.12 & 11 & 29844* & 11.70 & 29 & 24430 & 20.97 & 30 \\ 
         24359 & 22.19 & 30 & 30573 & 13.12 & 11 & \multicolumn{3}{c}{\textit{PSR J1958+2846}} & 23537 & 21.00 & 58 \\ 
         24371 & 22.19 & 29 & \multicolumn{3}{c}{\textit{PSR J1418$-$6058}} & 12149 & 0.00 & 10 & 24432 & 21.06 & 30 \\ 
         24361 & 22.21 & 30 & 2792 & 0.00 & 10 & 29841 & 13.61 & 9 & 24426 & 21.34 & 21 \\ 
         24374 & 22.22 & 28 & 2794 & 0.01 & 10 & 30699* & 13.62 & 10 & 26336 & 21.34 & 18 \\ 
         27256 & 22.23 & 20 & 7640 & 4.74 & 71 & 30700* & 13.62 & 10 &&&\\ 
    \hline \hline
    \end{tabular}
    \caption{\textit{Chandra} observations used in this analysis. Time since the first observation $\Delta t$ and exposure are listed. New observations collected for this survey are marked with an asterisk.}
    \label{tab:obs}
\end{table*}

\bibliography{bib}{}
\bibliographystyle{aasjournal}

\end{document}